\documentclass{amsart}

\usepackage{xcolor}
\usepackage[verbose]{hyperref}
\hypersetup{colorlinks=false,allbordercolors=blue,pdfborderstyle={/S/U/W 1}}
\usepackage{graphicx,color}
\usepackage{amsfonts}
\newtheorem{defn}{Definition}
\newtheorem{definition}{Definition}

\begin{document}

%\markboth{Authors' Names}{Instructions for typing manuscripts (Paper's Title)}

%\catchline{}{}{}{}{}
%
%%%%%%%%%%%%%%%%%%%%%%%%%%%%%%%%%%%%%%%%%%%%%%%%%%%%%%%%%%%%%%%%%%%%
\def\pd#1#2{\frac{\partial #1}{\partial#2}}
\def\<#1>{\langle#1\rangle}

 \def\rc#1{\begin{color}{red}#1\end{color}}
 \def\bc#1{\begin{color}{blue}#1\end{color}}
  \def\gc#1{\begin{color}{green}#1\end{color}}

\title{Alternative tangent and cotangent structures and their
physical applications}

\author{Jos\'e F. Cari\~nena}
\address{Departamento de F\'{\i}sica Te\'orica, Universidad de Zaragoza, Facultad de Ciencias-Campus San Francisco, 50009 Zaragoza (SPAIN)
\\
IUMA, Universidad de Zaragoza, Facultad de Ciencias-Campus San Francisco, 50009 Zaragoza (SPAIN) \\
}
\email{jfc@unizar.es}

\author{Jes\'us Clemente-Gallardo}
\address{Departamento de F\'{\i}sica Te\'orica, Universidad de Zaragoza, Facultad de Ciencias-Campus San Francisco, 50009 Zaragoza (SPAIN)
\\
Centro de Astropart\'{\i}culas y F\'{\i}sica de Altas Energ\'{\i}as (CAPA),  Universidad de Zaragoza, 50009 Zaragoza (SPAIN)
\\
}
\email{jcg@unizar.es}

\author{Giuseppe Marmo}

\address{INFN Sezione di Napoli and Dipartamento di Fisica "Ettore Pancini", Universit\`a Federico II di Napoli, Complesso Universitario di Monte Sant Angelo, Via Cintia, Napoli, I-80126 (ITALY) \\
}
\email{marmo@na.infn.it}

\maketitle

%\begin{history}
%\received{(Day Month Year)}
%\revised{(Day Month Year)}
%\accepted{(Day Month Year)}
%\published{(Day Month Year)}
%\comby{(Handling Editor)} % Communicated by optional
%\end{history}

\begin{abstract}
 The conditions under which a given manifold $M$ {\color{black}may} be given a tangent bundle or a cotangent bundle structure are analyzed. This is an important property arising in different 
 contexts. For instance, in the study of integrability of a given dynamics the existence of alternative compatible structures is very relevant,
 as  well as in the geometric approach to  Classical Mechanics.   
On the other hand in the quantum-to-classical transition, a Weyl system plays an important role for it provides (within the so-called Weyl-Wigner formalism) a description of quantum mechanics on a (symplectic) phase-space $M$. A Lagrangian subspace $Q\subset M$ of the (linear) phase space determines thus a maximal set of pairwise commuting unitary operators, which is used to parametrize the quantum states. As the choice of this maximal Abelian set of observables is not unique, the different choices make the phase space to become diffeomorphic to different cotangent bundles $T^* Q$ corresponding to different choices for the base manifold (and hence the fibers).
These motivating ideas are used to study   how to define alternative tangent and/or cotangent bundle structures on a phase space.
\end{abstract}



\section{Introduction: motivation to search for alternative structures}\label{sec:1}

The use of geometric methods in the study of  systems of differential equations and their applications in other branches of science and technology as classical and quantum dynamics \cite{AM78,CIMM,LM87}, 
control theory, biology, economy, etc., has been shown to be very useful and has received a lot of attention in recent decades.
This has shed light on many problems offering a new perspective  to analyse conceptual problems. When expressing all relevant relations in terms of tensorial objects we guarantee their 
independence of a particular choice of coordinates and we obtain the possibility of extending the theory to the case of  infinite-dimensional problems, 
maybe with the obstruction of topological properties. 

Integrability,  whatever its meaning be, can be studied by means of the analysis of symmetries and conserved quantities of the dynamical vector field, and the compatible
 structures defined by appropriate tensorial objects are particularly relevant in such a process.  The existence of 
alternative compatible structures and the fact of having a previous knowledge of alternative structures can be used to choose the most convenient one, or even to know other
 properties \cite{C24},  as first-integrals, for instance. In Newtonian mechanics, a dynamical system is described by a system of second-order differential equations on the configuration
  space $Q$. The existence of holonomic constraints  leads to consider that  the configuration space $Q$ is not an  affine space anymore but an $n$-dimensional differentiable manifold allowing to deal with the concept of differentiability. Then the {\color{black}system} of first-order differential equations {\color{black}is} replaced by {\color{black}a vector field} whose integral curves are  solutions  of  such a system.

   In order to describe Newtonian  dynamics, instead of dealing with {\color{black} a system} of second-order differential equations one prefers
    to describe it  by a system of first-order differential equations on a covering space of $Q$, usually either the tangent bundle 
 $\tau_Q: TQ\to Q$ or the cotangent bundle $\pi_Q:T^*Q\to Q$, in such a way that the solutions of this system of first-order differential equations reproduce, when projected 
 onto $Q$, all solutions of the original   equation. 
For instance, we can  write the system on the cotangent bundle, considering a possible Hamiltonian description in terms of a function $H\in C^\infty(T^*Q)$ and the natural  
Poisson bivector field $\Lambda$ on $T^*Q$ allowing the definition of a Poisson bracket $\{\cdot, \cdot \}$, given by $\{ f, g\}=\Lambda(df, dg)$, for $ f, g\in C^\infty(T^*Q)$.   
Then we can define the Hamiltonian vector field $\Gamma$ defined by $H$ as
$$
\Gamma(f)=\Lambda(df,dH),\quad \forall f\in C^ \infty(T^*Q),
$$
where $d$ represents the exterior differential on $M=T^*Q$. A diffeomorphism $\Phi:M\to M$ which is a symmetry for $\Gamma$ but not for $\Lambda$, when applied
 to the previous relation implies that
$$
\Gamma=\phi_*(\Gamma)=(\Phi_* \Lambda)(d ((\Phi^{-1})^* H)),
$$
i.e., the same dynamical vector field $\Gamma$  admits  an alternative Hamiltonian description on $M$, which, in general, will be endowed with a different cotangent bundle structure (see \cite{Giordano1993}), because $\Gamma$ is not only the Hamiltonian vector field of  $H$  with respect 
to $\Lambda$ but also the Hamiltonian vector of $H\circ \Phi^{-1}$ with respect to $\Phi_*\Lambda$.  Notice that when  $\Phi$ is a fiber-preserving diffeomorphism  we would
 go from one cotangent-bundle structure to an alternative one  with the same base manifold $Q$. Moreover, recall that  the Legendre transformation $\mathcal{F}L =TQ\to T^*Q$ can be introduced through the fiber-derivative, more explicitly as 
 \begin{equation} 
        \mathcal{F}L  (x,v) = (x, dL_x (v) \circ  \xi^v  )\ , \quad x=\tau(v).     \label{Legtransf}
 \end{equation}  
 Here for each   $v  \in  T_xQ$,   $\xi^v$  denotes the so called  vertical lift 
$\xi^v  : T_xQ \to T_v(TQ)$,  
{\color{black} defined by, }  
 \begin{equation} 
        \xi^v   (w) f = \frac d{dt } f(v+tw)_{ | t = 0},\ \forall f\in C^\infty({\color{black} U_v})\ ,        \label{vlift} 
         \end{equation} 
{\color{black}for $U_v$ an arbitrary neighborhood containing $v$, and }with local coordinate expression, if  $w  \in   T_qQ$  is $w = w^i  
\left(\partial/\partial {x^i}\right)_{|x}$,
\begin{equation} 
{\xi }{^v}(w)=w{^i} \left (\dfrac{\partial} {\partial v{^i}}\right)_{\mid (x,v)},       
\label{vlift2}
 \end{equation} 
and  $L_x: T_xQ \to \mathbb{R}$ is the 
function   $L_x(w) = L (x,w)$. Then if $v  \in   T_xQ$,   the map 
 $dL_x (v):T_v(T_xQ)\to\mathbb{R}$ is a linear map  and hence  $dL_x (v) \circ  \xi^v   : T_xQ \to \mathbb{R}$ is also linear  and  defines
an element of $T_x^*Q$. The Lagrangian function is said to be hyperregular when the map  $ \mathcal{F}L $ is a diffeomorphism.
   
    As the Legendre transformation relates the Lagrangian and Hamiltonian formalisms, when the dynamics admits  more than one (hyperregular) Lagrangian description with, say,  Lagrangians $L_1$ and $L_2$, the fiber preserving map
   obtained by composing the corresponding Legendre transformations $\mathcal{F}L_{2*}\circ (\mathcal{F}L_1^{-1})^*: T^*Q\to T^* Q$ would change the Liouville 
   1-form $\theta_0$  on $T^*Q$. This aspect is also relevant when we consider the quantum-to-classical transition to be commented later on. 
 
   A given tangent bundle structure gives an absolute meaning to `zero-velocity'  (the zero section of the bundle), clearly this is not appropriate when we move from one reference frame to a different one, indeed to take into account the `relative velocity' we are obliged to change the tangent bundle structure we consider on the same `configuration space'.

The reduction of many dynamical systems produces `reduced systems' on some carrier spaces (`reduced carrier spaces') which {\color{black} need not be}  tangent bundles or cotangent bundles
even if the starting dynamical system was defined on a  tangent  or cotangent bundle (see e.g. \cite{CGM07}). The possibility of introducing `conjugate variables' on 
the reduced carrier space  
or the  possibility of reading off the reduced system as a system of second-order differential equations (with a possible Lagrangian description) 
suggests us to investigate  when the reduced carrier space may be endowed with a tangent bundle structure or a cotangent bundle structure. Obviously there are some 
stringent conditions on the carrier space, for instance it should be  non-compact, because fibers should carry a linear space structure.  If there is a Poisson bracket
 the functions candidates to be position variables should be  a maximal set of pairwise commuting functions. 

Autonomous systems of $n$ second-order differential equations are   studied not only for its own 
mathematical interest, but also because they play a crucial r\^ole 
 in classical mechanics and in the spectral problem in Quantum Mechanics via the time-independent Schr\"odinger equation.  The usual way to proceed is to
  {\color{black}replace} the problem of a system of $n$ second-order differential equations {\color{black}with} a system of $2n$ first-order differential equations,
 or in geometric terms, to relate such systems of second-order differential equations {\color{black}with} a special kind of vector field $X$ on
  a tangent bundle $TQ$ of a $n$-dimensional manifold $Q$,   one of the class of second-order differential equation vector fields, 
 hereafter shortened as SODE vector fields. But the intrinsic property characterizing such special vector fields is written in terms of a (1,1)-tensor field $S$, which along with the dilation (or Liouville) vector field $\Delta$,
  completely identifies a tangent bundle structure \cite {Cr81,Cr83a, Morandi1990}.
 
  In a recent paper \cite{CMM23} the usefulness of alternative tangent bundle structures for giving a geometric interpretation of the Sundman transformation in the framework of SODE  vector fields
  has been shown.  Then one can try to  consider the Sundman transformation in the framework of SODE fields   in a  way similar to the first-order case \cite{CMM22}, 
i.e., by  changing the SODE vector field $X$ describing  the   similar
autonomous system to the vector field $f\, X$. The point however is 
that  in general $f\, X$ is not a SODE vector field anymore, and a more careful analysis is needed. Fortunately, the existence of alternative tangent   structures will be useful to solve this problem (see e.g. \cite{CMS19}).

In the context of Quantum Mechanics, we can also identify several examples where alternative cotangent bundle structures arise in a natural way. 
Consider a symplectic vector space $(V, \omega)$, where $\omega$ is assumed to be a translational invariant symplectic structure, a Hilbert space $\mathcal{H}$ and the corresponding 
group  of unitary operators $\mathbb{U}(\mathcal{H})$.  A \textsl{Weyl system} is defined as map  $W: V\to \mathbb{U}(\mathcal{H})$ such  that 
$$
W(v_1)W(v_2) W(v_1)^\dagger W(v_2)^\dagger=e^{i\omega(v_1, v_2)}\mathbb{I}, \quad \forall v_1, v_2\in V.
$$

{\color{black}By imposing the irreducibility of  the representation of the commutation relations, we may realize $\mathcal{H}$ as the space of square integrable functions on a Lagrangian subspace.}
A Weyl system allows us to quantize any classical function $f$ on $C^\infty(V)$ by considering it to be the symbol of the operator {\color{black}(see \cite{Gosson2014, Berge2022})}
$$
\hat A_f=N\int_V f^\omega (v) W(v) \omega^{\wedge n}(v),
$$
where $N$ represents a normalization factor, $\omega^{\wedge n}$ is  the symplectic volume element and $f^\omega (v)$ represents the \textit{symplectic Fourier transform} {\color{black}(see \cite{Gosson2014,Berge2022})} of the classical function $f\in C^\infty(V)$,  
$$
f^\omega (v)=N^{-1}\int_V f(v') e^{-i\omega(v,v')} \omega^{\wedge n}(v').
$$

It is also  know that in Quantum Mechanics, a basis of vectors of the abstract Hilbert space $\mathcal{H}$ may be selected by using a a complete set of commuting observables 
(hereafter  CSCO), made up from   a finite set of commuting self-adjoint linear operators on $\mathcal{H}$, say $\{ \hat O_1, \ldots, \hat O_n\}$, satisfying that each set of 
eigenvalues $ \{\lambda_1, \ldots, \lambda_n\}$ identifies a one-dimensional  subspace of common eigenvectors. Let us consider a case where the spectrum of each operator in this 
CSCO is isomorphic to $\mathbb{R}$ and hence  the set $\mathbb{R}^n$ represents the set of possible eigenvalues of the CSCO,  and $\mathcal{H}$ may be considered to be isomorphic to the completion of $(\mathcal{L}^2(\mathbb{R}^n), d\mu)$, where $d\mu$ represents the Lebesgue measure on that common spectrum. On this  Hilbert space, all operators in the 
 CSCO act as multiplicative operators. The  unitary operators $\{ e^{-i \hat O_1}, \ldots, e^{-i \hat O_n} \}$  pairwise commute and behave also as multiplicative
  operators on $\mathcal{H}$. Notice that each choice of CSCO determines a different unitary set. For instance, imagine that we consider a Hilbert space 
  $\mathcal{H}\approx \mathcal{L}^2(\mathbb{R}^2)$, and four different CSCO {\color{black}by means of position and momentum operators}:
$$ 
\{ \hat X_1, \hat X_2\}, \quad \{ \hat X_1, \hat P_2\}, \quad \{ \hat X_2, \hat P_1\}, \quad \{ \hat P_1, \hat P_2\}.
$$
The spectrum of each one of the  four operators is isomorphic to $\mathbb{R}$. Therefore, each choice determines a different (isomorphic) Hilbert space (with elements written 
as $\psi(x_1, x_2)$, $\psi(x_1, p_2)$, $\psi(p_1, x_2)$ or $\psi(p_1, p_2)$) and a set of unitary multiplicative operators
$$
\{ e^{-i\hat X_1}, e^{-i\hat X_2}\}, \quad \{ e^{-i\hat X_1}, e^{-i\hat P_2  }\}, \quad \{e^{-i \hat X_2}, e^{-i\hat P_1}\}, \quad \{ e^{-i\hat P_1}, e^{-i\hat P_2}\}.
$$
If we consider the subspace $Q\subset V$ generated by the pair of vectors $(e_1, e_2)$ in $V$ which under the Weyl mapping $W$ determines each pair  of unitary operators $(W(e_1), W(e_2))$, we realize that they generate different Lagrangian {\color{black}subspaces} of $V$. Indeed, they must be Lagrangian since, as the operators commute, 
$$
W(e_1)W(e_2)=W(e_2)W(e_1) \Longrightarrow \omega(e_1, e_2)=0.
$$
From this perspective, the space $(V, \omega)$ can be considered isomorphic with $T^*Q$ endowed with the natural symplectic structure of a cotangent bundle $\omega_0$. In this way, 
the vectors corresponding to the base manifold $Q=\mathrm{span}_\mathbb{R}(e_1, e_2)$,  are associated with multiplicative operators under the Weyl map. 

Therefore, the choice of a maximal set of pairwise commuting unitary operators to define the classical counterpart of our quantum model will identify a Lagrangian subspace on $V$. For each case, we will be able to define a \textit{different} symplectomorphism of $V$ with a cotangent bundle. Hence, it is natural to expect alternative cotangent bundle structures on a given symplectic vector space. We find that it is more economical to consider the problem from an algebraic point of view, and, from that algebraic analysis,  to identify the different geometrical structures encoding the bundle structures. As we are going to see, a very similar analysis allows us to define isomorphisms with alternative tangent bundle structures on $V$. 

A few references have considered these issues with some detail in the past, as \cite{Marmo1976, Caratu1976,Marmo1977} in the mid seventies. Later, much of the geometrical structures of the tangent and cotangent bundles were considered in \cite{Cr83a,Crampin1983a,CT} and extended in \cite{Thompson1987,Thompson1991}. A modern analysis of the ambiguities of the different structures was considered in \cite{CIMM}.

The structure of the paper is as follows. In Section  \ref{sec:2} we will recall the geometry of the tangent and the cotangent bundle structures as encoded in tensorial objects. These are the elements that we will try to identify in Section \ref{sec:3} when assuming that a manifold $M$ is given and we consider how to endow it with alternative structures. { \color{black} Section \ref{sec:4} presents a simple example to illustrate our construction. Finally, in Section \ref{sec:5} a summary of the main results of the paper and some future research lines are presented. \ref{sec:appendix} contains a brief discussion to provide a small complement to clarify further the meaning of the paper.}

\section{Intrinsic definition of vector bundle structures on a manifold $M$}
\label{sec:2}
\subsection{Tangent bundle geometry and tensor fields}

Let us recall now the main elements encoding the geometry of a tangent bundle, in order to be able to identify them later on. These properties have been extensively studied in 
\cite{Cr81,Cr83a} and we will try to present them in a more algebraic formulation where a manifold $M$ can be replaced by the  ring $\mathcal{F}=C^\infty(M)$ of
 real differentiable functions, because all relevant information on $M$ can be recovered from  $\mathcal{F}$.  The evaluation map ${\rm ev\,}:M\to {\rm Hom\,}(\mathcal{F},\mathbb{R})$  
 is bijective providing an identification of $M$ with $ {\rm Hom\,}(\mathcal{F},\mathbb{R})$. Observe that $\mathcal{F}$ is an associative and commutative real algebra 
 with a unital element.  Vector fields on $M$ are replaced by derivations of  $\mathcal{F}$, i.e., $\mathfrak{X}(M)$ by ${\rm Der\,} \mathcal{F}$, which is a real  Lie algebra, 
 and 1-forms on $M$ are 
 $\mathcal{F}$-linear maps from  ${\rm Der\,} \mathcal{F}$ to  $ \mathcal{F}$. Tensorial calculus on $M$ can be recovered from ${\rm Der\,} \mathcal{F}$  and $({\rm Der\,} \mathcal{F})^*$.  For instance, 
 the exterior algebra of forms is the  graded differential algebra built out from $({\rm Der\,} \mathcal{F})^*$  (see \cite{Nelson1967, LM90}).

In the  associative and commutative algebra  of functions $\mathcal{F}(TQ)$ of the tangent bundle $\tau_Q:TQ\to Q$, where $Q$ is an $n$--dimensional manifold,   we can identify the subalgebra of basic functions
$$
\mathcal{F}_Q=\{ f\in \mathcal{F}(TQ) \mid \exists f^Q\in \mathcal{F}(Q) \ {\rm such\ that}  \ f=\tau_Q^* (f^Q) \}.
$$
The $C^\infty(TQ)$-module of vertical vector fields on $TQ$, $\mathfrak{X}^{{\rm v}}(TQ)$, corresponds to those vector fields $Y$ on $TQ$ satisfying  $\mathcal{L}_Y f=0$, $\forall f\in \mathcal{F}_Q$:
$$
  \mathfrak{X}^{{\rm v}}(TQ)=\{ Y\in \mathfrak{X}(TQ)\mid  \mathcal{L}_Y f=0, \quad \forall f\in \mathcal{F}_Q\}.
$$
Notice that $\mathcal{F}(Q)$  and the set of vertical vector fields play a complementary role. As the distribution of vertical vector fields defines a fiber bundle structure on the manifold, the distribution is to be considered regular, and, therefore, also the subalgebra of basic functions will be called a regular subalgebra.

 Furthermore, each 1-form $\alpha\in\Omega^1(Q) $ defines  a fiberwise linear function  $\widehat \alpha$ on $TQ$, by $\widehat \alpha(q,v)=\alpha_q(v)$, 
with  local coordinate expression  in a tangent bundle chart on $\mathcal{U}$ obtained from a chart for  $Q$ on $U=\tau_Q(\mathcal{U})$ as follows: $\widehat \alpha(q,v)=v^i\, \alpha_i(q)$ when $\alpha=\alpha_i(q)\, dq^i$. In particular we can consider the case of $\alpha=df$ with 
$f\in \mathcal{F}(Q)$, and then the  fiberwise linear function on $TQ$, $\widehat{df}$,  is  $\widehat{df}(q,v)= v^i\dfrac{\partial f}{\partial q^i}$.
 
Given an  open set $U\subset Q$,  if $\{f^Q_1, \ldots, f^Q_n\}$  is a set of  functionally independent functions on $U$,  i.e., such that
  $df^Q_1\wedge \cdots \wedge df^Q_n \neq 0$, and as $\widehat{df_j}$ define fiberwise linear functions on $\tau^{-1}(U)$, 
  the set $\{f_1, \ldots, f_n, \widehat{d f_1}, \ldots , \widehat{df_n}\}$ can be considered as a set of independent functions on $\mathcal{U}=\tau^{-1}(U)$ and can be used to define
   a coordinate chart on  $\mathcal{U}$.

 It is also important to remark that in order to define the function $\widehat{df}$ as linear function on $TQ$ associated to the function $f$ on $Q$ we have made use of
  the vector bundle structure of  $TQ$, i.e.,   it involves the linear structure we have considered on the fibers of $TQ$.

%\begin{color}{green}
Recall that a  vector field  $Y\in\mathfrak{X}(TQ)$ is projectable on $X_Y \in\mathfrak{X}(Q)$   if  
$T\tau_Q \circ Y= X_Y\circ \tau_Q$, or equivalently, $\mathcal{L}_Y(\tau_Q^*f)=\tau_Q^*(\mathcal{L}_{X_Y}f)$, for all functions $f\in\mathcal{F}(Q)$, 
and that the $\mathbb{R}$-linear space 
  of projectable vector fields is a Lie algebra, with 
$X_{[Y_1,Y_2]} =[X_{Y_1}, X_{Y_2}]$. Vertical vector fields are those projectable on the zero vector field,   and moreover the set  $ \mathfrak{X}^{{\rm v}}(TQ)$ of vertical vector fields is  an ideal of the Lie subalgebra of projectable vector fields. The distribution generated by vertical vector fields  is involutive, and therefore integrable,   leaves are the fibers of the
 tangent bundle structure. 
%\end{color}

We conclude thus that the choice of the algebra of functions $\mathcal{F}_Q$ determines the base manifold, as the zero level set of $ (\widehat{d f_1}, \ldots , \widehat{df_n})$, and the fibers of the bundle structure. Let us see the remaining ingredients which transform that bundle structure on a tangent bundle one.

Also recall that the tangent bundle $TQ$, as any other vector bundle on $Q$, has associated a complete vector field $\Delta$, usually called  Liouville vector field, generator of dilations 
along the fibers, whose   local expression in usual tangent bundle coordinates is 
$   \Delta(q,v)=  v^i\partial/\partial {v^i}$. Moreover, in the tangent bundle case  there is also a natural (1,1)-tensor field, usually  called vertical endomorphism,  
which satisfies $\text{Im}\, S=\ker S$ and an 
   integrability condition, $N_S=0$   \cite{Cr81,Cr83a} (see also \cite{CMM23} and references therein), with a 
local coordinate expression  in usual tangent bundle coordinates
$ S=  (\partial/\partial {v^i})\otimes dq^i$.  
  The base manifold  $Q$ can be identified to the zero section for $\tau_Q$, while the  local expression of vertical vector fields on the tangent bundle $TQ$ are those  of the form 
  $D(q,v)= f^i(q,v)\partial/\partial {v^i}$, i.e., the vertical vectors are those of $\ker S$.
   In particular, the Liouville vector field $\Delta\in\mathfrak{X}(TQ)$ is a vertical vector field, i.e., $S(\Delta)=0$,  vanishing on the zero section and,  as the vertical endomorphism    $S$ is 
   homogeneous of degree minus one in velocities,  such that
   \begin{equation}
   \mathcal{L}_\Delta S=-S.\label{LDS} 
   \end{equation}

 {\color{black}
  The vector field $\Delta$ selects two subalgebras of vertical vector fields: 
 \begin{itemize}
  \item Fiberwise linear vector fields $Y$ such that $L_\Delta Y=0$. They are an $\mathcal{F}_Q$ module. 
  \item Fiberwise translational vector fields $T$ such that $L_\Delta T=-T$. They constitute an Abelian $\mathcal{F}_Q$ module and because of the completeness of $\Delta$, at each base point they generate the action of the translational Abelian vector group, whose orbits are the fibers of the tangent bundle.   With the help of $\Delta$, these fibers which are affine spaces become vector spaces.
 \end{itemize}

 Indeed},   $\Delta $ is the only vertical vector field on 
$TQ$ such that $\mathcal{L}_\Delta S=-S$ and vanishes on the zero section, i.e., $\Delta(q,0)=0$.

%\begin{color}{green}
The functions $\widehat \alpha $ defined on $TQ$ by 1-forms $\alpha$ on $Q$ satisfy $\mathcal{L}_\Delta \widehat \alpha= \widehat \alpha  $  and for any other function on $TQ$ such 
that $\mathcal{L}_\Delta f=f$, there exists a 1-form $\alpha$ on $Q$ such that $f= \widehat \alpha$, i.e., the functions $\widehat \alpha  $ span the linear subspace  of  functions which are fiber-wise linear.
 %\end{color}

This pair  of tensor fields, $\Delta$ and $S$, allows us to introduce the notion of second-order differential equation  vector fields as those $\Gamma\in \mathfrak{X}(TQ)$ 
such  that $S(\Gamma)=\Delta$,
whose local expressions  in local  tangent bundle coordinates  are
$\Gamma=v^k {\partial}/{\partial q^k}+ F^k(q,v) {\partial}/{\partial v^k}$. Thus,
the space of SODE  is an affine space whose vector space model is the space of vertical vector fields.
 
{\color{black}It follows easily that the set of partial linear structures that preserve $S$, i.e., such that $L_\Delta S=0$ constitute an affine space with $\Delta-\Delta'$ being a fiberwise translational vector field. Indeed we have $[\Delta, \Delta ']=-(\Delta'-\Delta)$ and $L_\Delta (\Delta'-\Delta)=-(\Delta'-\Delta)$.

It follows that t}he r\^ole  of tensor fields $\Delta$ and $S$ is different, {\color{black} one is responsible for the linear structure, while the other is responsible only for the affine structure. Indeed,  i}t was shown in \cite{CT} that the (1,1)-tensor field $S$ provides an affine bundle structure while $\Delta$ is responsible of the linear structure on the fibers.  As the characterisation  of SODE only depends on the pair  $(\Delta, S)$, similar  pairs $(\Delta', S')$ will give rise to different SODE vector fields. {\color{black} Instead of using the Liouville vector field $\Delta_Q$, one could use the approach, based on a regular action of the multiplicative monoid $(\mathbb{R}, \cdot)$  of reals
on {\color{black}$M$} as done in \cite{Janusz2009}}.

%\rc{Complete with something?}

There are several interesting {\color{black}relations, relating these tensor fields} and the ring of functions $\mathcal{F}_Q$. For instance:
\begin{itemize}
    \item Given a second order vector field $\Gamma$, the functions in the real linear space $\mathcal{L}_\Gamma \mathcal{F}_Q$ defined as
    $$
   \mathcal{L}_\Gamma \mathcal{F}_Q=\{ \mathcal{L}_\Gamma f\mid  f\in \mathcal{F}_Q \},
    $$
vanish identically on the zero section of the bundle $TQ$  since these functions are 
      linear functions    on the fiber coordinates, $ \mathcal{L}_\Gamma f=\widehat{df}$.
 \item For the same reason, any $f \in \mathcal{F}_Q$ satisfies 
    $$\mathcal{L}_\Delta (\mathcal{L}_\Gamma f)= \mathcal{L}_\Gamma f.$$
\end{itemize}

The affine nature of SODE, which we have mentiond earlier, implies the following properties:

\begin{itemize}
 \item For any pair $\Gamma, \Gamma'$ of SODE vector fields,                                                                                                   $$\mathcal{L}_\Gamma f= \widehat{df}=  \mathcal{L}_{\Gamma'}  f, \quad \forall f\in \mathcal{F}_Q.
    $$
    \item The commutator with  $\Delta$ of any SODE vector field $\Gamma$ differs from $\Gamma$ by   a vertical vector field:
    $$
[\Delta, \Gamma]-\Gamma \in \mathfrak{X}^{{\rm v}}(TQ).
    $$
Hence, for any $f\in \mathcal{F}_Q$,
    $$
\mathcal{L}_{[\Delta, \Gamma]-\Gamma} f=0.
    $$
    
    \item Again, as  the action of any SODE vector field $\Gamma$ on a basic function defines a fiber-wise linear function,  this implies that
    $$ S^* (d (\mathcal{L}_\Gamma f))=df \quad \text{ and }  \quad S^* (df)=0,   \quad \forall f\in \mathcal{F}_Q.$$
    This property implies that from sufficient number of functions $f\in \mathcal{F}_Q$, we can generate pairs  $(df, d(\mathcal{L}_\Gamma f))$ of 1-forms on $TQ$ which are sufficient to define a basis of one forms on $TQ$. Then, previous relation is enough to define the endomorphism $S$.
\end{itemize}

{\color{black}Remark: Most of the stated properties, except for the completeness of the partial linear structure, hold true also for an open submanifold of $TQ$ containing the zero section. The completeness of $\Delta$ implies that each fiber is an orbit of the Abelian vector group with dimension equal to the dimension of $Q$.}

These properties will guide  us later on to identify alternative tangent bundle structures on the same $2n$--dimensional manifold $M$, once we build the geometrical tools.

\subsection{Cotangent bundle geometry and tensor fields}

Let us proceed in a similar way with the case of the cotangent bundle.  Notice that the definition of the basic functions and the linear structure on the fibers on a given cotangent  $T^*Q$  
is completely valid also in this case. Let us recall how an appropriate additional 1-form  $\theta$ defined on the manifold, determines the cotangent bundle structure on it.

 If $Q$ is a given $n$-dimensional manifold, its cotangent bundle $\pi_Q:T^*Q\to Q$ is a 
 vector bundle  endowed with a canonical 1-form $\theta_0$ such that $\omega_0=-d\theta_0$ is a symplectic form. Such a 1-form is defined as 
  \begin{equation}     \theta_{0 \alpha}(Y)=\alpha (\pi_{Q*\alpha}Y)\,, \quad\forall  \alpha\in T^*Q,\ 
 Y \in T_\alpha(T^*Q)\ ,       \label{defLiou1form}
  \end{equation} 
i.e.,  such that $  \theta_{0 \alpha}= (\pi_Q) ^*_{\alpha}\alpha$.
Moreover, as for every vector bundle, the base manifold can be identified with the zero section, and its fibers are endowed with a vector field $\Delta$ which  is the infinitesimal generator of 
dilations. In this case $\Delta$ is the vector field in the kernel of $\theta_0$ associated with the 1-form $\theta_0$ by the symplectic form, by
$$
i(\Delta) \omega_0 =-\theta_0.
$$
 
{\color{black}The vector field $\Delta$ is a complete vector field and hence we may proceed as we did in the tangent bundle case: we can build an Abelian vector group whose orbits are the fibers of the (now) contangent bundle. }

Given a coordinate chart $(U,\varphi)$ of $Q$, we can define a chart on $\mathcal{U}^*=\pi_{Q}^{-1}(U) $ in a similar way as we did 
with $\mathcal{U}=\tau_{Q}^{-1}(U) $: The linear space of forms at a point of $Q$ is the dual of the tangent space at such a
 point and we consider the dual basis to the one considered in the tangent space. More explicitly, the fiber coordinates at such a point of the covector 
$\alpha$ are
  \begin{equation} 
  p_i(\alpha)=\alpha\left(\frac{\partial }{ \partial q^i}_{\mid\pi(\alpha)}\right). \label{defpi}
  \end{equation} 
  This means that we may consider vector fields on the base manifold as fiber-wise linear functions on the cotangent bundle.

In such a coordinate system the local expressions of $\Delta$ and $\theta_0$ are
$$\Delta =p_i\, \frac{\partial }{ \partial p_i},\qquad  \theta_{0}=p_i\, dq^i.
$$
Note that $\omega_0=-d\theta_0$ is given by  $\omega_0=dq^i\wedge dp_i$,  which  is an exact  symplectic form, i.e., such that $\omega_0^{\wedge n}\ne 0$.  

%\begin{color}{green}
 Similarly  to what happens in the tangent bundle $\tau_Q:TQ\to Q$ with one-forms,  in the dual case of the cotangent bundle  $\pi_Q:T^*Q\to Q$,  each vector field on the base manifold  
 $X\in\mathfrak{X}(Q)$ has an associated function $\widehat X$ on  the cotangent bundle, as follows: $\widehat{X}(\alpha)=\alpha(X(\pi_Q(\alpha))$.  If the  coordinate expression of the 
 vector field is  $X=X^i(q)\partial/\partial q^i$,  then  $\widehat{X}(x,p)=p_i\,X^i(q)$, i.e., the functions $\widehat{X}$ are functions linear along the fibers of  $T^*Q$.
 
 The concept of projectable vector field can also be defined:  a  vector field  $Y\in\mathfrak{X}(T^*Q)$ is projectable on $X_Y \in\mathfrak{X}(Q)$   if  
$T\pi_Q \circ Y= X_Y\circ \pi_Q$, or equivalently, $\mathcal{L}_Y(\pi_Q^*f)=\pi_Q^*(\mathcal{L}_{X_Y}f)$, for all functions $f\in\mathcal{F}(Q)$, 
and that the $\mathbb{R}$-linear space   of projectable vector fields is a Lie algebra, with 
$X_{[Y_1,Y_2]} =[X_{Y_1}, X_{Y_2}]$. Vertical vector fields are those projectable on the zero vector field on $Q$,   and moreover the set  $ \mathfrak{X}^{{\rm v}}(T^*Q)$ of vertical vector fields is 
an ideal of the Lie algebra $ \mathfrak{X}^{\mathrm{proj}}(T^*Q)$. The distribution generated by vertical vector fields  is involutive, and therefore integrable,  whose leaves are the fibers of the
 tangent bundle structure, which are Lagrangian submanifolds. 
%\end{color}

Each fiber is a leaf of the vertical foliation defined by the $n$ one-forms $dq^1, \ldots, dq^n$ (notice that there is a certain abuse of notation, as they are basic forms on the bundle, i.e., the pullback of 1-forms on $Q$)  or by  a multiple by a nonvanishing function $f$ on $T^*Q$ of the $n$-form 
$\eta=dq^1\wedge\cdots \wedge dq^n$. Notice that $\theta_0\wedge \eta=0$. As $\eta$ is only defined up to product by a nonvanishing  function, even if $d\eta=0$, in general 
$d(f\eta)=df\wedge \eta=(df/f)\wedge (f\eta)$.  

Note also that as $i(\Delta)\omega_0=-\theta_0$, this implies that $\mathcal{L}_\Delta\omega_0=\omega_0$.
These remarks will be useful and used when we shall consider the existence of alternative cotangent bundle structures: An almost symplectic structure on a  $2n$-dimensional manifold $M$ endowed with a partial linear structure is given by 
a 2-form $\omega$ such that  $\omega^{\wedge n}\ne 0$ and a Lagrangian distribution $\mathcal{L}$, that is to say,  $\mathcal{L}$ is an $n$-dimensional distribution such that 
when evaluating $\omega$ on any two vectors of $\mathcal{L}$ the result is zero. The  almost cotangent structure is integrable when $\omega$ is closed, $d\omega=0$,
and the  distribution $\mathcal{L}$ is integrable, i.e., involutive. More information on these structures can be found in  \cite{Thompson1987,Thompson1991}.

\section{Constructing alternative bundle structures on a manifold}
\label{sec:3}
Let us consider now the following problem: imagine we are given a $2n$-dimensional manifold $M$. We want now to search for a vector bundle structure on $M$, if it exists, and define tangent or cotangent bundle structures. And consider what is the freedom we have to do these choices. We are going to see that, inspired by the properties discussed above, we can associate the structures to a certain subalgebra of functions, a vector field, and some additional ingredients. Let us analyze it.

We will consider three different steps:
\begin{itemize}
    \item First, we will consider what do we need to define on $M$ a fiber bundle structure. In order to do this, we will use the analog to the algebra of basic functions that we saw in  previous section.
    \item Then, we will introduce a linear structure on the fibers of the bundle with the help of a dilation vector field, making it a vector bundle.
    \item Finally, we will consider how it is possible to define a tangent or a cotangent bundle structure, by introducing either a $(1,1)$-tensor field $S$ or a 1-form $\theta$. Both objects are related with the $(1,1)$ identity tensor field on $Q$ which in local coordinates reads $\mathrm{I}=dq^k\otimes \frac{\partial}{\partial q^k}$. Indeed,  when defining $S$ we consider the vertical lift to the bundle of the vector field component  {\color{black}in} $S=dq^k\otimes \frac{\partial}{\partial v^k}$,
       and when defining $\theta$ we consider {\color{black}the} vector field component as a fiber-wise linear function on the dual bundle $\theta=p_kdq^k$.
\end{itemize}

But, first of all, let us remind how the geometric elements which are used in Mechanics can be encoded in algebraic terms.

\subsection{Encoding geometric properties in algebraic terms}
%\begin{color}{green} 

  Many of the properties of the associative and commutative algebra $\mathcal{F}(M)$ can be extended to more general associative and commutative algebras
  $\mathcal{F}$
 over a ring, for instance for its subalgebras (see e.g. \cite{geroch,CCLM17,LM90}).
 Within this framework, several geometric properties can be represented in simple terms.

 We can consider, for instance, the case of derivations to represent vector fields.
 Recall that a derivation on $\mathcal{F}$  consists of a mapping $\xi :\mathcal{F} \to \mathcal{F} $  such that, for all pair of elements $f,g\in\mathcal{F} $: 
 
 \begin{itemize}
  \item[i)]$\xi (f + g) = \xi (f) + \xi (g)$
  \item[ii)] $\xi (fg)=f\,\xi (g) + \xi (f)\, g$, and
  \item[iii)]  if $f\in  \mathcal{R}$, then $\xi (f) = 0$
 \end{itemize}

 When $\mathcal{F}= \mathcal{F}(M)$ the collection ${\rm Der\,} \mathcal{F}$   of derivations (satisfying suitable locality requirements, see \cite{Peetre60})  is just the collection of smooth contravariant 
vector fields on the manifold $M$. Note that ${\rm Der\,} \mathcal{F}$ is also a $ \mathcal{F}$-module, because if $\xi$  and
$ \eta $ are derivations, and $g\in\mathcal{F}$, we can define a new derivation, $(\xi  + g\,\eta )$, by 
$$(\xi + g\,\eta ) (f) = \xi(f) + g\,\eta (f),\quad \forall g\in  \mathcal{F}, \, \forall \xi,\eta\in {\rm Der\,} \mathcal{F} .$$

 Hence, ${\rm Der\,}\mathcal{F}$ is a module over the ring
 $\mathcal{F}$. Furthermore, if $\xi$  and $\eta$  are derivations, then 
 $$(\Xi_\xi \eta )(f) = \xi (\eta (f)) - \eta (\xi (f))$$ 
 defines a new derivation $\Xi_\xi \eta$, the Lie bracket of $\xi$  and $\eta $. This defines a Lie algebra structure on ${\rm Der\,}\mathcal{F}$.
 
 Let  $({\rm Der\,}\mathcal{F})^*$
denote the dual module of ${\rm Der\,}\mathcal{F} $, i.e., an element $\mu \in ({\rm Der\,}\mathcal{F})^*$  associates with each
 $\xi\in{\rm Der\,}\mathcal{F}$ an element $\mu(\xi)$  of $\mathcal{F}$ in a  $\mathcal{F}$--linear way: 
 $$\mu(\xi  + g\, \eta ) = \mu(\xi ) + g\,\mu(\eta ).$$ 
 When $\mathcal{F}= \mathcal{F}(M)$
  the module $({\rm Der\,}\mathcal{F})^*$  is   the set  of smooth covariant vector fields, i.e., 1-forms, on the manifold $M$,  the elements of  $\Omega^1(M)$.
  
In this language, a metric can be represented as an isomorphism $g$ from the module ${\rm Der\,}\mathcal{F}$  to the module  $({\rm Der\,}\mathcal{F})^*$ which is symmetric, i.e., such that 
$$g(\xi , \eta ) = g(\eta , \xi ),$$ for all $\xi , \eta \in  {\rm Der\,}\mathcal{F}$,
 where $g(\xi , \eta )$ is given by  $g(\xi , \eta ) = (g(\xi )) (\eta )$.
Analogously, a covariant tensor field of rank $n$ is a multilinear mapping $\alpha : {\rm Der\,}\mathcal{F}\times \cdots\times  {\rm Der\,}\mathcal{F}\to  \mathcal{F}$ ($n$ factors). For  contravariant 
tensor fields, use can be made  of the  isomorphism $g$   between  ${\rm Der\,}\mathcal{F}$ and $({\rm Der\,}\mathcal{F})^*$.
%\end{color}

\subsection{Fiber bundle structure} 

In order to define a fiber bundle structure on $M$ we are going to need to define analogs of the ingredients that we considered for the tangent and cotangent bundles in the previous section. In particular, we are going to need a subalgebra $\mathcal{F}_Q$ of the algebra $\mathcal{F}(M)$ of smooth functions of $M$, which contains enough information to build the bundle structure. Remember that, equivalently, we may define the distribution defined by the set of vector fields which preserve and annihilate $\mathcal{F}_Q$, which we denote as $\mathfrak{X}^{\mathrm{v}}(M)$ and call \textit{vertical vector fields}:
$$
\mathfrak{X}^{\mathrm{v}}(M)=\{ Y\in \mathfrak{X}(M) \; | \; Y(f)=0, \quad \forall f \in \mathcal{F}_Q \}.
$$
It is immediate to verify that $\mathfrak{X}^{\mathrm{v}}(M)$ defines a Lie subalgebra of $\mathfrak{X}(M)$ and an ideal in $\mathrm{Der}\mathcal{F}_Q$, the set of derivations of $\mathcal{F}_Q$. We can then consider the short exact sequence of Lie algebras (or also of $\mathcal{F}_Q$ modules)

\begin{equation}
    \label{eq:shortexact}
0\rightarrow \mathfrak{X}^\mathrm{v}(M) \rightarrow \mathrm{\rm Der\,}(\mathcal{F}_Q) \rightarrow \mathfrak{X}^Q\rightarrow 0.
\end{equation}
Notice that  $\mathfrak{X}^Q$ is the set of equivalence classes  (with respect to $\mathfrak{X}^{\mathrm{v}}(M)$) of vector fields, that are derivations of $\mathcal{F}_Q$. 
Furthermore  $\mathfrak{X}^\mathrm{v}(M)$ defines an involutive distribution of $\mathcal{F}(M)$, and therefore a foliation for $M$, where the set $\mathfrak{X}^\mathrm{v}(M)$ determines the tangent spaces to the leaves.  Thus, each leaf is a maximal integral submanifold of the vertical distribution.

In the following, we will consider a subalgebra $\mathcal{F}_Q$ which is a regular subalgebra of $\mathcal{F}(M)$, where the notion of regularity is defined as:
\begin{definition}
Let $M$ be a differentiable manifold. We shall say that a subalgebra $\mathcal{F}_Q\subset \mathcal{F}(M)$ is regular if the foliation defined by its distribution of vertical vector fields is {\color{black} (strictly) simple (see \cite{Moerdijk2003} and \cite{GGKM2024} for a brief discussion)} and therefore its space of leaves can be given a structure of differentiable manifold, which will be denoted as $Q$. {\color{black}Furthermore, we will ask the subalgebra $\mathcal{F}_Q\subset \mathcal{F}(M)$ to be full, i.e.,  for any $Y\in \mathfrak{X}^v(M)$, $L_Y f=0$ implies that $f\in \mathcal{F}_Q$.
The associated bundle structure $\pi^M:M\to Q$  makes $\mathcal{F}_Q=\pi^{M*}(\mathcal{F}(Q))$. }
\end{definition}

{\color{black}We now make the requirement that at each point of $Q$, the algebra of vertical vector fields contains an Abelian subalgebra of complete vector fields giving rise to an action of an Abelian vector group of dimension $k=n-r$.}

Consider then a subalgebra $\mathcal{F}_Q$ which is regular {\color{black}and full} according to the definition above. 
For each point $m\in M$ there is a neighborhood $\mathcal{U}$ such that the restriction  $${\mathcal{F}_Q}_{|\mathcal{U}}=\{ f_{|\mathcal{U}}\; | \;  f\in \mathcal{F}_Q\}$$ contains $r$  functionally   independent functions 
 $\{f_1, \ldots , f_r\} $, i.e., such that   
$(df_1 \wedge \cdots \wedge df_r)(m)\ne 0$  for all points $m\in \mathcal{U}$,  and  that 
 for any function $f \in  \mathcal{F}_Q$, 
$(df\wedge df_1 \wedge \cdots \wedge df_r)(m)=0$,  $\forall  m\in \mathcal{U}$. In this case, we will say that $\mathcal{F}_Q$, and therefore the manifold $Q$, is $r$--dimensional, while the fiber of $M$, determined by the set of vertical vector fields $\mathfrak{X}^{\mathrm{v}}(M)$, will be $k=n-r$ dimensional.

Having introduced a bundle structure on $M$, we can also consider the concept of projectable vector field. A vector field  $Y\in\mathfrak{X}(M)$ is projectable on $X_Y \in\mathfrak{X}(Q)$ (w.r.t. $\pi^M:M\to Q$) if  
$T\pi^M \circ Y= X_Y\circ \pi^M$, or equivalently, $\mathcal{L}_Y(\pi^{M*}f)=\pi^{M*}(\mathcal{L}_{X_Y}f)$, for all functions $f\in \mathcal{F}(Q)$. The $\mathbb{R}$-linear space 
  of projectable vector fields is a Lie algebra, with 
$X_{[Y_1,Y_2]} =[X_{Y_1}, X_{Y_2}]$ and, moreover, the subalgebra of vertical vector fields is an ideal of such subalgebra, because if $Y$ is an arbitrary projectable vector field and 
$Z\in \mathfrak{X}^{{\rm v}}(M)$, then  for each $f\in \mathcal{F}_Q$, 
$$  \mathcal{L}_{[Z,Y]}(\pi^{M*}f))=\mathcal{L}_Z (\mathcal{L}_Y(\pi^{M*}f))- \mathcal{L}_Y (\mathcal{L}_Z(\pi^{M*}f))= \mathcal{L}_Z (\pi^{M*}(\mathcal{L}_{X_Y}f))=0.
$$
Consequently, the following sequence of Lie algebras is exact:
$$
0\rightarrow \mathfrak{X}^\mathrm{v}(M) \rightarrow \mathfrak{X}^{{\rm proj}}(M) \rightarrow \mathfrak{X}(Q)\rightarrow 0.
$$

Notice the similarities with exact sequence (\ref{eq:shortexact}), where projectable vector fields on the bundle $M$ take now the place of $\mathrm{\rm Der\,} \mathcal{F}_Q$. Indeed, they are equivalent, as we can see. First, notice that this last sequence can also be considered a sequence of $\mathcal{F}_Q$ modules, once we notice that  $\mathcal{F}_Q=\pi^{M*}(\mathcal{F}(Q))$. Indeed, for all $f_1,f_2,g\in \mathcal{F}(Q)$,

\begin{multline*}
X_{[\pi^{M*}(f_1)Y_1,\pi^{M*}(f_2)Y_2]} g=\pi \left([\pi^{M*}(f_1)Y_1,\pi^{M*}(f_2)Y_2] \pi^{M*}(g) \right )= \\\pi^M \left(
\pi^{M*}(f_1)Y_1\pi^*(f_2)Y_2 \pi^{M*}(g)-\pi^{M*}(f_2)Y_2\pi^{M*}(f_1)Y_1 \pi^{M*}(g) \right )=\\
=\left (f_1X_{Y_1} f_2 X_{Y_2} g - f_2X_{Y_2} f_1 X_{Y_1} g \right )=
[f_1X_{Y_1}, f_2X_{Y_2}] g .
\end{multline*}

It can also be proved that the modules $\mathfrak{X}^{{\rm proj}}(M)$ and $\mathrm{\rm Der\,} \mathcal{F}_Q$ coincide when quotiented by the vertical vector fields.
It is clear that  projectability condition $\mathcal{L}_Y(\pi^{M*}f)=\pi^{M*}(\mathcal{L}_{X_Y}f)$, implies that $\mathfrak{X}^{{\rm proj}}(M) \subset \mathrm{\rm Der\,} \mathcal{F}_Q$, because 
if $Y$ is a projectable vector field $Y\in \mathfrak{X}^{{\rm proj}}(M)$, as  
$Y(\pi^{M*} f)=\pi^{M*} ( X_Y (f)) ,\ \forall f\in \mathcal{F}(Q) $, we see that $Y\in \mathrm{\rm Der\,} \mathcal{F}_Q$. Moreover, if $Y^{\rm v}$ is a vertical vector field, then $Y+Y^{\rm v}$ produces the same derivation in $\mathcal{F}_Q$.  Previous relation allows to define $X_Y$  for a given $Y$, because it holds true for any function $f \in \mathcal{F}_Q$.

Notice that, having a bundle structure on $M$, we can consider also the concept of connection on that bundle. In order to define a connection on the manifold $M$, we must consider a 
decomposition of the tangent bundle $TM$ in such a way that the tangent space at a point $m\in M$ can be written as a direct sum of the linear space $V(m)$ of vertical vectors at that 
point and a suitable complementary space $H(m)$, the  horizontal subspace. Any idempotent endomorphism $C$  of $\mathfrak{X}(M)$ (i.e., any idempotent $(1,1)$--tensor field on 
$M$) which is the identity on $\mathfrak{X}^{\mathrm{v}}(M)$ defines such a decomposition. Choosing $\mathbb{I}-C$ we can select a representative in each equivalence
 class of $\mathfrak{X}^Q\simeq \mathfrak{X}(Q)$, which defines the (horizontal) {supplementary} subspace at that point.  Analogously, we can see the connection as a 
 mapping $\iota:\mathfrak{X}^Q \to \mathrm{\rm Der\,}\mathcal{F}_Q$, which selects {a representative} of the equivalence class of derivations of $\mathcal{F}_Q$, 
 realised as a vector field on $M$. This allows to write $\mathrm{\rm Der\,} \mathcal{F}_Q\sim \mathfrak{X}^\mathrm{v}(M)\oplus \iota(\mathfrak{X}^Q)$.

 {\color{black}With the various assumptions we have made, we have now a fiber bundle having as typical fiber an orbit of an Abelian vector group of dimension $k$.}

%\gc{This should be used to proved the rivial locality of $Q$ over $Q$.}

 \subsection{Vector bundle structure}
The choice of a regular subalgebra $\mathcal{F}_Q$ {\color{black}and the requirement that the corresponding set of vertical vector fields contains a maximal Abelian subalgebra of complete vecgtor fields} can be used to  define a fiber bundle structure on an $n$--dimensional manifold $M$. But, what is it furthermore necessary to make it a vector bundle?

The  {linear} structure on the fibers requires of a {\color{black}complete} vector field $\Delta_Q$ {\color{black}(what we have called a partial linear structure)} on $M$  which makes the fibers of the foliation  a linear space. We address the interested reader to \cite{CIMM} for a more detailed presentation of these concepts.

Let $\Delta $ be a complete {\color{black}partial linear structure} on a differentiable manifold $M$. {\color{black}It} can be seen as a derivation of the commutative algebra  $\mathcal{F}(M)$. It gives rise to a gradation 
on the ring $\mathcal{F}(M)$ of differentiable functions, with $  \mathcal{F}^{(k)}_\Delta(M)$ being the set of homogeneous functions of degree $k$ w.r.t. $\Delta$:
$$  \mathcal{F}^{(k)}_\Delta(M)=\{f\in \mathcal{F}(M)\mid   \mathcal{L}_\Delta f=k\, f\}, \quad k\in \mathbb{N}.
$$
  Linearity and Leibniz rule property for derivations allow us to check that $  \mathcal{F}^{(0)}(M)$ is a subalgebra of $  \mathcal{F} (M)$ while {\color{black}$  \mathcal{F}^{(k)}(M)$} is not a subalgebra but only a $  \mathcal{F}^{(0)}(M)$-module. As we have seen  before, if we consider the algebra $\mathcal{F}^{(0)}(M)$ to define the base manifold $Q$, then   `vertical vector fields', defined by   $Y\in\mathfrak{X}(M)$ such that $ \mathcal{L}_Yf=0$, 
  $\forall f\in  \mathcal{F}^{(0)}(M)$, 
  span an involutive, therefore integrable,  distribution, and the functions of  $  \mathcal{F}^{(0)}(M)$ can be seen as functions on the space of leaves, i.e., functions on $Q$.      
  
   Let us assume that the differentials  of functions of $  \mathcal{F}^{(0)}(M)$  and $  \mathcal{F}^{(1)}(M)$ span the set of forms $\Omega^1(M)$. We can also assume that 
   the algebra $  \mathcal{F}^{(0)}(M)$ is locally finitely generated by $f_1^{(0)},\ldots,f_r^{(0)}$,  and that the restrictions of $\Delta $ to the level sets of $f_1^{(0)},\ldots,f_r^{(0)}$,
   define linear structures on the leaves. Similarly we assume that $  \mathcal{F}^{(1)}(M)$ is locally finitely generated by $f_1^{(1)},\ldots,f_k^{(1)}$, where $n=r+k$. {\color{black}These assumptions qualify $\Delta$ as a partial linear structure}:
   
   \begin{defn}  
   We say that a complete vector field $\Delta$ in a manifold $M$ is a partial linear structures when  $\mathcal{F}^{(0)}(M)$  and $  \mathcal{F}^{(1)}(M)$ are finitely generated projective modules, such that  
   $  d\mathcal{F}^{(0)}(M)$  and $ d \mathcal{F}^{(1)}(M)$ span $\Omega^1(M)$  as an $\mathcal{F}^{(0)}(M)$-module, and, moreover, the set of critical points  of $\Delta$ is a submanifold  
   whose commutative algebra of functions is isomorphic to $\mathcal{F}^{(0)}(M)$ (for details see \cite{CIMM}). 
  \end{defn} 
  
  {Thus, fiber-wise linear functions (with respect to such a partial linear structure $\Delta$) are those $f\in \mathcal{F}(M)$ satisfying that $ \mathcal{L}_{\Delta}f=f$, that is, the elements of $\mathcal{F}_\Delta^{(1)}$.}  
    If we denote the functions $f_i^{(0)}$ by $x^i$ and the functions $f_\alpha^{(1)}$ by  $y^\alpha$ the coordinate expression of $\Delta$ will be 
  
$$\Delta=y^\alpha\,\frac{\partial}{\partial y^\alpha}.$$ 

Having introduced a partial linear structure, we can also consider {\color{black}a subalgebra of the} projectable vector fields.
Let $\mathfrak{X}^{(1)}(M)$ denote 
$$\mathfrak{X}^{(1)}(M)=\{X\in \mathfrak{X}(M)\mid [X,\Delta]= \mathcal{L}_{X}\Delta=0\}.$$
This is a real Lie subalgebra of $ \mathfrak{X}(M)$ because it is a $\mathbb{R}$-linear space and such that if $X_1,X_2\in \mathfrak{X}^{(1)}(M)$, then, the Jacobi identity of vector fields leads to
$$[[X_1,X_2],\Delta]=-[[X_2,\Delta],X_1]-[[\Delta,X_1],X_2]=0.
$$

These  vector fields $X\in \mathfrak{X}^{(1)}(M)$   preserve the $\mathbb{R}$-linear subspace $\mathcal{F}_\Delta^{(0)}$ of basic functions, i.e., they satisfy 
$ \mathcal{L}_{X}(\mathcal{F}_\Delta^{(0)})\subset\mathcal{F}_\Delta^{(0)}$, 
because from the relation 
$$ \mathcal{L}_{\Delta}( \mathcal{L}_{X} f)- \mathcal{L}_{X}( \mathcal{L}_{\Delta} f)=\mathcal{L}_{[\Delta,X]}f,\quad f\in \mathcal{F}(M),
$$ we see that 
 if $f\in \mathcal{F}_\Delta^{(0)}$  and $X\in \mathfrak{X}^{(1)}(M)$, i.e., $[X,\Delta ]=0$, then 
$$ \mathcal{L}_{\Delta}( \mathcal{L}_{X} f)= \mathcal{L}_{X}( \mathcal{L}_{\Delta} f)=0,$$
and then $ \mathcal{L}_{X}(\mathcal{F}_\Delta^{(0)})\subset\mathcal{F}_\Delta^{(0)}$.  But note that, as indicated before, this is nothing but {\color{black}a particular case of }  the projectability condition.

Conversely, if $X\in \mathfrak{X}(M)$  is such that  $ \mathcal{L}_{X}(\mathcal{F}_\Delta^{(0)})\subset\mathcal{F}_\Delta^{(0)}$, i.e., $X$ is projectable, 
then for each function $f\in \mathcal{F}_\Delta^{(0)}$, as $\mathcal{L}_{\Delta} f=0$ and then $ \mathcal{L}_{\Delta}( \mathcal{L}_{X} f)=0$,  we see that 
$\mathcal{L}_{[\Delta,X]}f=0$, which is a weaker condition than $ [X,\Delta]=0$.

This is the reason why vector field $X\in\mathfrak{X}^{(1)}(M)$ are said to be fiberwise linear when $[\Delta,X]=0$. Their  coordinate expressions in the above mentioned coordinates are  
 $$X=f^i(x^1,\ldots,x^r)\pd{}{x^i}+a^\alpha_{\ \beta} (x) y^\beta\pd{}{y^\alpha}.
 $$

Let us consider again the bundle associated with the algebra $\mathcal{F}_Q$. Clearly,  the construction of  a vector bundle structure requires considering a complete vector field $\Delta_Q$ on $M$ satisfying:
\begin{itemize}
    \item The set of critical  points, i.e.,  $p\in M$ where $\Delta_Q$ vanishes, defines a submanifold $Q$ whose algebra of functions is isomorphic to  $\mathcal{F}_Q$.
    \item The eigenvalue problem
    $$
 \mathcal{L}_{\Delta_Q}f=0, \quad f\in \mathcal{F}(M)
    $$
    defines an algebra  $\mathcal{F}^{(0)}(M)$ of solutions isomorphic to $\mathcal{F}_Q$.
    
     \item  Homogeneous functions of degree one w.r.t. $\Delta_Q$, i.e., those $f\in \mathcal{F}(M)$  satisfying that  
     $$
  \mathcal{L}_{\Delta_Q}f=f, \quad f\in \mathcal{F}(M),
     $$
correspond to a $\mathcal{F}_Q $-module $\mathcal{F}^1_Q$.
 
  \item $  d\mathcal{F}_Q$  and $ d \mathcal{F}^{1}_Q$ span $\Omega^1(M)$  as an $\mathcal{F}^{(0)}(M)$-module
\end{itemize}

Endowed with such a partial linear structure, the bundle structure of the manifold $M$ becomes a vector bundle on the manifold $Q$, since $\Delta_Q$ defines a linear structure on each fiber of $\pi^M:M\to Q$. The set of $\pi^M$-vertical vector fields $\mathfrak{X}^\mathrm{v}(M)$ becomes then  a $\mathcal{F}_Q $-module  and the corresponding set of leaves of the associated foliation the base of the vector bundle. Remark however that  the commutative algebra $\mathcal{F}_Q$ only determines {\color{black}an open neighborhood of the zero section} and not the {\color{black}full} vector bundle structure.
For instance, the same base manifold $Q$ is determined by all partial linear structures which are affinely related.  An important particular case to be analyzed in next Subsection  corresponds to the case $r=k=n$.

\subsection{Tangent and cotangent structures}
\subsubsection{Tangent bundle structure}
If we are now interested in tangent bundle structure we need  $r=k=n$ in the $2n$-dimensional manifold $M$ endowed with a partial linear structure $\Delta_Q${, defining the vector bundle structure $\pi^M:M\to Q$ through the algebra $\mathcal{F}_Q$ defining $Q$}
  and we need 
as an additional   further ingredient  for the vector bundle to become a tangent bundle an appropriate  $(1,1)$-tensor field. 

Then, a vector field $X\in \mathrm{\rm Der\,}\mathcal{F}(M)$ is said to be a \textsl{second-order differential equation vector field} (SODE) {\color{black}with respect to $\mathcal{F}_Q$,} if
\begin{equation}
\mathcal{L}_{\Delta_Q} \left( \mathcal{L}_X  (\pi^{M*}f)\right)= \mathcal{L}_X (\pi^{M*} f) , \quad \forall f\in \mathcal{F}(Q),
\label{eq:SODE}
\end{equation}
{\color{black}along with $L_{\Delta_Q}f=0$.}

Notice that such a vector field $X$ {\color{black}(it if exists)} allows us to generate  $\mathcal{F}_Q^1$ from 
$\mathcal{F}_Q$ as
$\mathcal{F}_Q^1 = \{\mathcal{L}_X  (\pi^{M*}f) \mid  f \in  \mathcal{F}(Q)\}$,
and    we can define  the {linear space} of 1-forms on $M$ by  using the set of functions $f \in  \mathcal{F}(Q)$  
 as
$d\mathcal{F}_Q= {\rm span\,}\{\pi^{M*} (df) \mid  f \in  \mathcal{F}(Q)\}$
and $d\mathcal{F}_Q^1 = {\rm span\,}\{\mathcal{L}_X (\pi^{M*} df)\mid  f \in  \mathcal{F}(Q)\}$. {\color{black}Therefore, relation \eqref{eq:SODE} actually determines $\Delta_Q$ when such an $X$ exists.}

 Again, because of the regularity  conditions above and the partial-linear structure, we {\color{black} may  define} a basis of 1-forms on every neighborhood $\mathcal{U}\subset M$ from the elements in $\mathcal{F}_Q$ and $\mathcal{F}_Q^{1}$. We also have the analogous basis for the vector fields. {\color{black}We assume that there exists a SODE as \eqref{eq:SODE} and a partial linear structure $\Delta_Q$. Then, we can define a $(1,1)$--tensor $S_Q$
  by setting
  $$
S_Q(L_X \pi_M^* df)=\pi_M^* df, \qquad S_Q (\pi_M^*df)=0.
  $$
  This (1,1) tensor satisfies} the compatibility conditions with the linear structure $\Delta_Q$ to define a tangent bundle structure.

  The local expression of the tensor $S_Q$ in the coordinates   $\{x^i, \ldots,x^n,y^1,\ldots, y^n\}$  introduced above becomes:
  $$
S_Q=d x^k\otimes \frac{\partial}{\partial y^k}.
  $$
  
  By taking the Lie derivative with respect to $\Delta_Q$, as $ \mathcal{L}_{\Delta_Q} (dx^k)=0$ and $ \mathcal{L}_{\Delta_Q} (\partial/\partial y^k)=-\partial/\partial y^k$, we find that that
 $$
  \mathcal{L}_{\Delta_Q} S_Q=-S_Q.
 $$

 Given this pair of tensors $(\Delta_Q, S_Q)$, we can define a tangent bundle structure $M=TQ$. In such a case, a vector field $X$ will be a {SODE} vector field if
 $$
S_Q (X)=\Delta_Q.
 $$

\subsubsection{Cotangent bundle structure}
So far we have proved that a suitable regular subalgebra $\mathcal{F}_Q$ on a manifold $M$ and a partial linear structure $\Delta_Q$ 
{are} able to define a vector bundle structure $\pi^M:M\to Q$ on $M$.
Then the concepts of vertical vector field and  semibasic form are well defined.
If the dimension of the base happens to be half the dimension of $M$, the vector bundle can be endowed with a tangent bundle structure, {\color{black} if it exists a SODE. This is done by}  introducing a suitable $(1,1)$--tensor $S_Q$.  Fixing the pair $(\Delta_Q, S_Q)$, $M$ becomes diffeomorphic to $TQ$.  We would like to study now what are the additional requirements to be considered on $M$ for it to become diffeomorhic to a cotangent bundle.  
 
We know that the usual cotangent bundle structure is encoded in the choice of the semibasic  (i.e., vanishing on any vertical vector field) 1-form $\theta_0 \in \Omega^1(M)$.
Hence,  let $\theta_Q$ be a semibasic 1-form on $M$, such that $d\theta_Q$ is of rank $2n$   and the compatibility condition reads just
$i(\Delta_Q) d \theta_Q= \theta_Q$, that implies that $\Delta_Q\in \ker \theta_Q$.
This  leads to  the condition, 
$$ \mathcal{L}_{\Delta_Q} \theta_Q=\theta_Q,$$ 
and consequently, 
$$ \mathcal{L}_{\Delta_Q} \omega_Q=\omega_Q.$$ 
These can be considered the defining conditions of the possible cotangent bundle structures on $M$, once the {\color{black} completeness condition for $\Delta_Q$ is satisfied.}

Finding a general solution for the equation above may be hard, in general. Nonetheless, we must remember that, given a vector bundle $\pi^M:M\to Q$, we can always consider the corresponding dual bundle, which will be diffeomorphic to the original one (but not canonically diffeomorphic). The dual bundle, that we can represent as $\pi^{M ^*}:M^*\to Q$, will be a vector bundle, having as fiber the dual space to the fiber of $M$ {\color{black}at each point of $Q$,} and a partial linear structure $\Delta_Q^*$ that will be related with the one on $M$ as:
\begin{itemize}
  \item The set of critical points of $\Delta^*_Q$ (i.e., the base manifold of the bundle $M^*$) must coincide with the set of critical points of $\Delta_Q$.
  \item The 0-th order functions $\mathcal{F}^{0*}$ must coincide with $\mathcal{F}_Q^0$
  \item The fiber-wise linear functions $\mathcal{F}^{1*}$ must define a $\mathcal{F}^0$--module isomorphic to the fiber-wise linear module $\mathcal{F}_Q^1$. This follows from the consideration that one-forms on $Q$ are isomorphic with vector fields on $Q$, and one-forms define fiber-wise linear functions on $TQ$ while vector fields define fiber-wise linear functions on $T^*Q$ 
\end{itemize}
By construction, as we know that, having fixed a pair of tensors $(\Delta_Q, S_Q)$, $M$ is diffeomorphic to the tangent bundle $TQ$, its dual bundle $M^*$ must be diffeomorphic to $T^*Q$. All we need is a way to define a particular diffeomorphism. 
Thus, one of the possible definitions of a cotangent bundle structure considers the introduction of a Riemannian structure $g$  on the base manifold $Q$ and the corresponding geodesic Lagrangian.

The Riemannian tensor allows us to define a mapping
$$
\hat g:M\to M^*,
$$
which maps the fiber-wise linear functions of $M$ on the fiber-wise linear functions of $M^*$, i.e.,
$$
\mathcal{F}^{1*}=\{ (\hat g^{-1})^* f\mid  f\in \mathcal{F}_Q^1\},
$$
where $\hat g^{-1*}$ represents the pullback of the bundle isomorphism $\hat g^{-1}:M^*\to M$. As we have already noticed, the metric tensor creates an isomorphism between one-forms and vector fields on $Q$ which corresponds to this isomorphism between fiber-wise linear funtions.

More specifically, we introduce the following function $\mathcal{L}_g\in \mathcal{F}(M)$:
$$
\mathcal{L}_g(m)=\frac 12 g_{\pi^M(m)}(m,m),
$$
where $ g_{\pi^M(m)}$ represents the value of the tensor field $g$ at $\pi^M(m)\in Q$.
 If we use the local trivialization of the tangent bundle structure and the coordinate system  $(x^k, v^k)$ seen above, the local expression of the function becomes
$$
\mathcal{L}_g(x, y)=\frac 12 g_{jk}(x)v^j v^k; 
$$
where $\quad x=(x^1, \ldots, x^n); \quad v=(v^1,\ldots, v^n)$.

If we consider now the $S_Q$ tensor above, and define the differential action on $\mathcal{L}_g$, we can define a one-form on $M$:
$$
\theta^{g}=d \mathcal{L}_g\circ S_Q.
$$
In this way, we recover a one-form on $M$ which will help us to define a diffeomorphism with a cotangent bundle. 
In the basis above, the coordinate expression of $\theta^g$ becomes

$$
\theta^g(x,v)=\frac{\partial \mathcal{L}_g}{\partial v^k}  dx^k.
$$
By construction, this one-form satisfies:
$$
i_{\Delta_Q} d\theta^g=\theta^g,
$$
and therefore
$$
 \mathcal{L}_{\Delta_Q} \theta^g=\theta^g.
$$

The coordinate expression of the Legendre transformation associated with the Lagrangian $\mathcal{L}_g$ using the local trivialization of the bundle in the basis above becomes
$$
\mathcal{FL}_g( x,v)=(x, \hat g(v)).
$$
As $\hat g$ is a one-to-one map from the set of fiber-wise linear functions of $M$ to the set of fiber-wise linear functions of $M^*$, we can define a natural basis $(x,p)$ on $M^*$ from the basis corresponding to $(x,v)$ on $M$.  Hence, in local coordinates 
$$
p_k:= \hat g(v)_k=\frac{\partial \mathcal{L}_g}{\partial v^k},
$$
what allows us to write:
$$
\mathcal{FL}_g( x,v)=(x, p).
$$
Analogously, 
$$
\mathcal{FL}^{-1}_g( x,p)=(x, {\hat g}^{-1}(p))=(x,v).
$$
Then, the one-form $\theta\in \Omega^1(M^*)$ defined as 
$$
\theta=(\mathcal{FL}^{-1})^* \theta^g,
$$
defines a cotangent bundle structure on $M^*$.
By construction, the coordinate expression of $\theta$ is particularly simple:
$$
\theta(x,p)=p_k dx^k.
$$
As we see, the mapping we defined has identified the canonical one-form $\theta^0$ on the cotangent bundle $T^*Q$ with the object defined on the dual bundle $M^*$.

A few comments are in order:
\begin{itemize}
   \item Notice that what we just argued is simple to understand since, after choosing $\Delta_Q$ and $S_Q$, $M$ becomes  diffeomorphic to the tangent bundle $TQ$. Then the manifold $M^*$ can be considered diffeomorphic to $T^*Q$.  Nonetheless, as we may consider any Riemannian structure, any one of them will define an alternative Legendre transformation and therefore a one-form $\theta$ for the cotangent bundle $M^*$. By composing the Legendre maps associated with them we build alternative contangent bundle structures with respect to a reference one.
The appropriate composition of Legendre maps will define what are known as \textit{fouling transformations} (see  \cite{CurrieSaletan66}).

   \item From the expression of $\theta$  we already see that adding to the Lagrangian function any `total time derivative' of a function on the base manifold will shift the zero section of the bundle $M^*\to Q$, therefore providing a different vector (and hence cotangent) bundle structure on $M^*$.
   Notice that this freedom justifies choosing the Legendre transformation as diffeomorphism, and not using directly the metric tensor, which would not offer this choice in such a simple (and well-known) form.  
   \item Furthermore, notice that, besides defining alternative structures on $M^*$ for a fixed description of $M$ as a tangent bundle,  alternative choices of tangent bundle structures are also possible by choosing different pairs $(\Delta_Q, S_Q)$. Through the Lagrangians $\mathcal{L}_g$, each one of those choices would also yield alternative cotangent bundle structures on $M^*$. 
   
   \item Finally, it is important to remark that our choice fixing the tangent bundle structure first and determining from it the cotangent bundle one is by no means necessary. We presented it in this form for the sake of simplicity as the different mappings are simpler to write. Nonetheless, we might also have determined first a cotangent bundle structure on the vector bundle $M$,  and then choose a Hamiltonian function to define a Legendre transformation and determine from it alternative tangent bundle structures (i.e., tensors $S_Q$) on the dual bundle. The choice of alternative cotangent bundle structures would yield then alternative tangent bundle ones.

\end{itemize}

{\color{black}\section{A simple example}
\label{sec:4}
Consider a manifold $M=\mathbb{R}^2$ and the set of canonical coordinates $\{x_1, x_2\}$ with a harmonic oscillator dynamics defined on it:
$$
\dot x_1= x_2, \qquad \dot x_2=-x_1.
$$
Consider now the change of coordinates:
$$
q=f(x_1^2+x_2^2)x_1, \qquad v=f(x_1^2+x_2^2)x_2;
$$
for a differentiable nowhere vanishing function $f:\mathbb{R}\to \mathbb{R}$.  Dynamics in the new coordinates reads:
$$
\dot q=v; \qquad \dot v=-q.
$$

We can build the tangent bundle structure on these two different set of coordinates, which are related by a non-linear transformation which, clearly, does not preserve the bundle structure.

Let us consider the subalgebra $\mathcal{F}_Q$ to be the algebra of functions of this variable $q$. Hence, 
a vector field $Y=A(x_1,x_2)\frac{\partial}{\partial x_1}+B(x_1,x_2)\frac{\partial}{\partial x_2} \in \mathfrak{X}(\mathbb{R}^2)$ will be a vertical vector field for this function if 
$$
Y(q)=0\Rightarrow A(x_1, x_2)(2x_1^2f'+f)+ B(x_1, x_2)2x_2x_1f'=0
$$
Hence the vertical distribution becomes:
$$
 \mathfrak{X}^v(\mathbb{R}^2)= \left \{ -2x_2x_1 f'B\frac{\partial}{\partial x_1} + (2x_1^2f'+f)B\frac{\partial}{\partial x_2}; \qquad  B\in C^\infty(\mathbb{R}^2) \right \}.
$$
This distribution can also be written as:
$$
\left \{ C\frac{\partial}{\partial v}; \qquad C \in C^\infty(\mathbb{R}^2)  \right \}.
$$

Clearly, we can consider a partial linear structure defined on $\mathbb{R}^2$ by the vector field
$$
\Delta^b_Q= v\frac{\partial}{\partial v},
$$
which makes $\mathbb{R}^2$ a vector bundle with base defined by the linear subspace generated by the coordinate $q$. It is straightforward to verify that this structure is not compatible with the partial linear structure defined by the vector field which, in the original coordinates, reads:
$$
\Delta^a_Q= x_2\frac{\partial}{\partial x_2},
$$
and the corresponding vector bundle structure.

Both vector bundles can be endowed with tangent structures by introducing two tensor fields:
$$
S_Q^b=dq\otimes \frac{\partial}{\partial v},
$$
and
$$
S_Q^a=dx_1\otimes \frac{\partial}{\partial x_2},
$$
where the two structures are globally defined. Thus, we can define two alternative tangent bundle structures by choosing:
\begin{itemize}
  \item The associative algebra of functions of the variable $x_1$, which has a vertical distribution the vector fields proportional to $\partial_{x_2}$. On this bundle, we define a linear structure with the vector field $\Delta_Q^a$ and a tangent bundle structure adding the tensor field $S_Q^a$.
  \item The algebra of functions of the variable $q$, which has a vertical distribution the vector fields proportional to $\partial_{v}$. On this bundle, we define a linear structure with the vector field $\Delta_Q^b$ and a tangent bundle structure adding the tensor field $S_Q^bº$.
\end{itemize}

If we select any metric on $\mathcal{R}$, we can define an isomorphism with the corresponding cotangent bundle structures. Choosing a trivial metric equal to the identity, we can define two cotangent bundle structures
$$
\theta_a=p_1\, dx_1, \qquad \theta_b=p \, dq
$$
where the dual coordinates are defined as
$$
p_1=\frac{\partial \mathcal{L}_a}{\partial x_1}; \qquad
p=\frac{\partial \mathcal{L}_b}{\partial q},
$$
and the Lagrangian functions are
$$
\mathcal{L}_a=\frac 12 x_1^2; \qquad \mathcal{L}_b=\frac 12 q^2.
$$

It is straightforward to generalize this case to a $2n$--dimensional framework where we associate together pairs of variables $(x_1, x_2, \ldots, x_{2n-1}, x_{2n})\to (q_1, v_1, \ldots, q_n,v_n)$. From the point of view of applications, these alternative structures become extremely useful in the context of $f$--deformed oscillators and $f$--coherent states, for systems where the frequency of oscillations depends on the energy (see \cite{Manko1996}). The choice of these alternative structures allows a straightforward integration of the dynamics.

}

\section{Conclusions and outlook}
\label{sec:5}

%\begin{color}{black} {\bc To be rewritten}

In this paper we have studied the problem of the definition of alternative tangent and cotangent bundle structures on a given manifold $M$. We have done this by considering a subalgebra $\mathcal{F}_Q$ of the algebra of functions of the manifold or, equivalently, an integrable distribution of vector fields which preserve it and which we call vertical. We have imposed a regularity condition on $\mathcal{F}_Q$, requiring that the foliation defined by the vertical vector fields {\color{black}defines a fiber bundle}, {\color{black}i.e., there exists a maximal Abelian subalgebra of complete vertical vector fields generating the action of an Abelian vector group.} 
This condition implies that the space of leaves of the foliation becomes a differentiable manifold $Q$ and the projection map
  applying to each element in its fibre  defines a fiber bundle. {\color{black}A} partial-linear structure $\Delta_Q$ defined on this bundle, {\color{black}turns} the fiber into a linear space, {\color{black}and } $M$ becomes a vector bundle.  

{\color{black}When the dimensions of the algebra and the vertical distribution coincide (and fibers are vector spaces), it defines a tangent bundle structure on M by
requiring the existence of a second order vector field.}
 %it is possible to introduce a $(1,1)$--tensor field $S_Q$ that, with $\Delta_Q$, %defines a tangent bundle structure on $M$.
  Finally, if we consider a Riemannian structure on the base manifold $Q$, we can consider on $M$ the hyper-regular Lagrangian function defined by the metric tensor. Using the corresponding Legendre transformation, we can define a diffeomorphism of $M$ with a cotangent bundle on the base manifold $Q$. Similar ideas  were introduced by Libermann in \cite{Libermann2000} and discussed later in \cite{Tulczyjew2008} to elaborate on the ideas introduced in \cite{Alekseevsky1994}. 

From an abstract point of view, given a manifold $Q$ the tangent or the cotangent bundle structures are uniquely defined. However, from the point of view of physical interpretation (identification of  physical variables) it may be useful and convenient to consider alternative tangent or cotangent bundle structures due to different identification of variables, e.g., going from one reference frame to another in relativity or the introduction of a gauge when a partcle moves in some external magnetic potential.

Different choices of the different algebraic and geometric objects considered, as the subalgebra $\mathcal{F}_Q$ (or, equivalently, the vertical distribution), the partial-linear structure $\Delta_Q$, the vertical endomorphism $S_Q$ and the metric tensor $g$ on the base manifold, would lead to alternative tangent/cotangent bundle structures.  For instance,  diffeomorphisms which take from one subalgebra to
another will automatically define a new geometrical structure on the
same manifold. This type of approach generalizes the results presented in \cite{deFilippo89}.

From the point of view of applications, these different alternative structures on the manifold may offer interesting alternative classical limits for quantum systems. By changing the `Lebesgue measure',
it may be used to turn quantum differential operators into Hermitian ones (see \cite{Avanzo2005}).  At the classical level, it may be used to provide Hamiltonian descriptions for reparametrized vector fields, or `regularized' vector fields. We can also consider the definition of  `non-Noether' constants of the motion, and so on. These problems  will be considered in future works.

\vspace{1cm}
\underline{\textbf{Acknowledgements}}
This paper is dedicated to the memory of Miguel Carlos Mu\~noz-Lecanda, one of the pioneers and main figures of the geometric approach to mechanics and control systems in Spain. For your friendship, your support and all the knowledge that you so generously shared with us, thank you Miguel. 

The authors acknowledge partial ﬁnancial support of Grants PID2021-125515NB-C22 (JFC) and PID2021-123251NB-I00 (JCG) funded by MCIN/AEI /10.13039/501100011033 and by the European Union, and of Grant E48-23R funded by Government of Aragón.
 GM is a member of the Gruppo Nazionale di Fisica Matematica
(INDAM), Italy.

{\color{black}
\appendix
\section{Dynamics and geometrical structures}
\label{sec:appendix}

Prompted by the interesting comments of the referees, we have decided to add {\color{black}a} few remarks to the main text of the paper in the form of {\color{black}an} appendix, to avoid changing the structure of the paper.

First we would like to stress our belief that one of the most important achievements of last century mathematics was the realization that mathematical structures may be studied “per se”, i.e., ”abstractly”, without any representation or realization.
Thereafter their use in physics requires the “identification of physical variables”,  and then abstract results possess a physical interpretation. The same mathematical structure may have completely different physical interpretation. See the example provided in \cite{Esposito2014}, in section 4.7, example 1.
There it is shown that the same Poisson manifold may describe the electron-monopole system or a massless particle with helicity, depending on the identification of physical variables. The description of physical systems requires not only abstract mathematical models but also a specific realization with an identification of physical variables.

To illustrate the relations of abstract structures and physical interpretations, our favorite {\color{black}example} is provided by the three-dimensional Lie algebra called Heisenberg-Weyl algebra. In terms of a basis, say $\{E_0,E_1,E_2\}$, {\color{black}the }commutation relations are:
$$
[E_1,E_2]=E_0, \quad [E_0, E_1]=0=[E_0,E_2].
$$
We can construct the tensorial algebra on it and the enveloping algebra at the abstract level. A realization in terms of “canonical conjugate variables” provides a Poisson bracket and all classical mechanics in terms of Hamiltonian description, a realization of the same algebra in the Cartan’s dual form in terms of one-forms provides contact manifolds and contact mechanics, say
$$
\{ pdq+ds, dp, dq\}.
$$
In terms of $q$ and $i\hbar\frac{\partial}{\partial q}$, provides a description of Schrödinger wave mechanics, in terms of position and momentum operators acting on some complex infinite dimensional Hilbert space provides a description of Dirac-von Neumann quantum mechanics, and a realization in terms of creation and annihilation operators provides a description in terms of Fock spaces of quantum field theory.
Our point of view concerning mathematical structures appearing in physical theories, expressed by some of us in previous collaborations (see \cite{CIMM,Marmo1985}), is that dynamical evolution has a {\color{black}fundamental} primary role and other structures have a “derived “role”.
In the spirit of Klein’s program, in our approach the relevant group of transformations is the group of transformations which preserve the dynamical vector field. If other geometrical structures, represented by tensorial quantities, appear in the description of our dynamical vector field, applying diffeomorphisms which preserve the dynamical vector field but do not
preserve the additional tensor fields we go from one geometrical structure to an alternative one; both of them compatible with the given dynamical. Bi-Hamiltonian or bi-Lagrangian, or bi-unitary structures are well known examples of this situation. Perhaps the existence of alternative tangent bundle or cotangent bundle structures is less known. These latter aspects are taken into account in the present paper.
If we consider a second order vector field and “reparametrize” it by means of a factor function (as $1+H^2$, $H$ being  a constant of the motion), we obtain clearly an infinitesimal {\color{black}symmetry} which does not respect the fibre bundle structure neither of the tangent bundle nor of the cotangent bundle. Thus, the description of the two bundles, tangent and cotangent, by means of tensorial quantities opens the way to alternative structures.
}

%\bibliographystyle{ws-gm}
%\bibliography{/home/jesus/Zotero/jesus/jesus-bibtex.bib}
%\printbibliography
\end{document}